\documentclass[english,prb,twocolumn,notitlepage,superscriptaddress] {revtex4-1}
\pdfoutput=1
\usepackage{graphicx}
\usepackage{amssymb}
\usepackage{amsmath}
\usepackage{setspace}
\usepackage{epstopdf}
\usepackage{multirow}

\begin{document}

\title{Valley blockade and multielectron spin-valley Kondo effect in silicon}
\author{A. Crippa}
\email{alessandro.crippa@mdm.imm.cnr.it}
\affiliation{Dipartimento di Scienza dei Materiali, Universit\`a di Milano Bicocca, Via Cozzi 53, 20125 Milano, Italy}
\affiliation{Laboratorio MDM, CNR-IMM, Via C. Olivetti 2, 20864 Agrate Brianza (MB), Italy}
\author{M. L. V. Tagliaferri}
\author{D. Rotta}
\affiliation{Dipartimento di Scienza dei Materiali, Universit\`a di Milano Bicocca, Via Cozzi 53, 20125 Milano, Italy}
\affiliation{Laboratorio MDM, CNR-IMM, Via C. Olivetti 2, 20864 Agrate Brianza (MB), Italy}
\author{M. De Michielis}
\affiliation{Laboratorio MDM, CNR-IMM, Via C. Olivetti 2, 20864 Agrate Brianza (MB), Italy}
\author{G. Mazzeo}
\author{M. Fanciulli}
\affiliation{Dipartimento di Scienza dei Materiali, Universit\`a di Milano Bicocca, Via Cozzi 53, 20125 Milano, Italy}
\affiliation{Laboratorio MDM, CNR-IMM, Via C. Olivetti 2, 20864 Agrate Brianza (MB), Italy}
\author{R. Wacquez}
\author{M. Vinet}
\affiliation{CEA-LETI-MINATEC CEA-Grenoble, 17 Rue des Martyrs 38054 Grenoble, France}
\author{E. Prati}
\email{enrico.prati@cnr.it}
\affiliation{Istituto di Fotonica e Nanotecnologia, CNR, Piazza Leonardo da Vinci 32, 20133 Milano, Italy}
\affiliation{Laboratorio MDM, CNR-IMM, Via C. Olivetti 2, 20864 Agrate Brianza (MB), Italy}

%\begin{document}

\begin{abstract}
We report on the valley blockade and the multielectron Kondo effect generated by an impurity atom in a silicon nano field effect device. According to the spin-valley nature of tunnelling processes, and consistently with those allowed by the valley blockade regime, the manifestation of Kondo effect obeys to the periodicity 4 of the electron filling sequence typical of silicon emerging at occupation $N=1, 2, 3$. The spin-valley Kondo effect emerges under different kinds of screening depending on the electron filling. By exploiting the valley blockade regime, valley index conservation in the Kondo SU(4) is deduced without the employment of an external magnetic field. Microwave irradiation suppresses the Kondo effect at occupancies up to three electrons.
\end{abstract}

\maketitle

\section{INTRODUCTION}
The control of individual electrons and impurity atoms in silicon nano Field Effect Transistors (FETs) has deep impact on the field of valleytronics \cite{behnia_valleytronics}, which consists of exploiting the orbital part of the electron wave function as additional degree of freedom with respect to more conventional charge and spin states. 
Valley-related effects have been lately reported, from the valley filling sequence in silicon quantum dots (Si-QDs) \cite{Morello_spinValley, Xiao2010parallel, DeMichielis_APEX} and lifetime-enhanced transport as indirect observation of spin and valley blockade \cite{Rogge_LET}, to a tunable valley Kondo effect in a As atom at single electron filling ($N=1$) \cite{Rogge_kondoPRL}. Valley-based qubits \cite{Culcer2012valley} and pure valley blockade \cite{Prati_valley} have been proposed in analogy to spin qubits and Pauli spin blockade respectively. 
The direct experimental observation of the valley blockade, namely the suppression of transport determined by the orthogonality of orbital degrees in a reservoir and at an impurity site, was till now unaddressed. Valley quantum numbers may also influence higher-order tunneling processes involving two or more particles, as in the case of the Kondo effect.
The simplest Kondo theory for III-V semiconductor quantum dots accounts for the emergence of Kondo effect only with uncoupled spins in the dot \cite{goldhaber_nat_98, Cronenwett, goldhaber_prl_98}.
Generalized to the case of Si-QDs and group-V donors, the Kondo physics predicts that both spin and valley indices of a confined electron can be screened by a Fermi sea through virtual spin-valley flip processes, allowing for highly symmetric SU(4) Kondo states \cite{HE-kondo}. \\
We already demonstrated single/few donors based silicon transistors operability \cite{Mazzeo, Prati_NatNano, leti_switching} and microwave irradiation effects in single atom based devices for spectroscopy of excited states at zero bias \cite{Prati_MW} as well as in single interface defects \cite{Prati_PLA, Prati_JAP}.\\
Here we investigate a system of a P atom strongly coupled to the leads of a silicon ultra-scaled tri-gate FET (Fig. \ref{fig:dev}a) \cite{Afsid}. Peculiarities of valley dependent quantum transport arise both in sequential tunneling processes and in Kondo-like cotunneling events. 
The amount of valley mixing in contact-impurity tunnelings gives rise to different kinds of Kondo screening from $N=1$ to $N=3$ at the relatively high temperature of 4.2 K. In agreement with theory, standard Coulomb blockade is observed at $N=4$ where Kondo effect is forbidden. 
 The fourfold degeneracy is usually revealed by splitting the orbital screening from the spin screening with a magnetic field \cite{Rogge_kondoNL, Rogge_kondoPRL, deFranceschi_kondoCNT}. Differently, in our experiment the valley blockade of first-order transport ($N=0 \to 1 \to 0$) influences the Kondo processes within the $N=1$ Coulomb diamond: it selects donor-lead transitions between identical valley states, which beautifully confirms the selection rule at the base of the SU(4) symmetry.
The Kondo perturbed regime is obtained at the base temperature at filling up to 3 electrons and it is suppressed by means of microwave irradiation. \\
The paper is organized as follows. In Sec. II the device and the experimental methods are described. In Sec. III the experimental results are presented and discussed. The Appendix connects the spin-valley states mentioned in Sec. III with the established formalism of Ref. \citenum{HE-PRB}. In Supplemental Material details about device electrical characterization and data analysis procedure are reported.\\
\section{DEVICE AND METHODS}
\begin{figure}
\centering
	\includegraphics[width=\columnwidth]{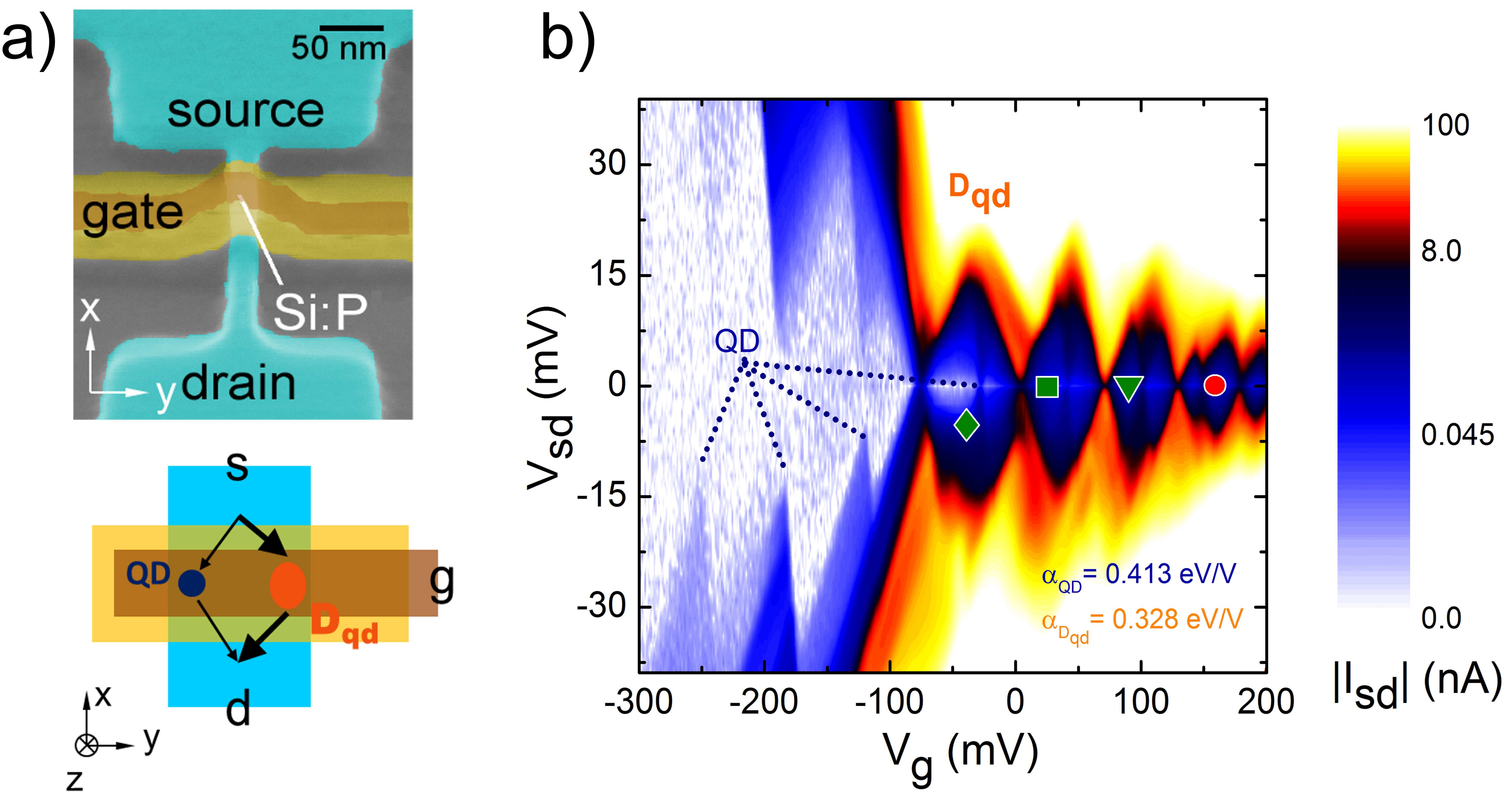} 
	\caption{(Color online) (a) Scanning Electron Micrograph image of a tri-gate FET identical to measured device. Below the sketch of the active region with the two conductive paths. Brown denotes the gate contact whereas Si$_3$N$_4$ spacers around the gate are yellow. Heavily-doped regions of reservoirs are highlighted in blue. (b) From the stability diagram two Coulomb blockade patterns appear, each with its own lever-arm factor $\alpha$. Red-orange (high current) diamonds are attributed to the donor $\text{D}_{\text{qd}}$ strongly coupled with leads. Blue (low current) diamonds are relative to a weakly coupled quantum dot.}
	\label{fig:dev}
\end{figure}

The tri-gate FET is fabricated from fully depleted silicon-on-insulator (FDSOI) technology. The silicon device layer is phosphorous implanted ($10^{18}$ $\text{cm}^{-3}$) and then etched to define the nanowire channel (Fig. \ref{fig:dev}a). The gate is then formed on the three sides of the narrow channel with a 5 nm thermal silicon dioxide of isolation. Silicon nitride (Si$_3$N$_4$) spacers around the gate protect the active region from Arsenic implantation of highly doped (metal-like) source-drain electrodes. Nominal channel width of the tested device is 50 nm, gate length is 20 nm and channel is 8 nm thick. A dc voltage $V_g$ sets the gate potential to tune the number of confined electrons, a bias $V_{sd}$ controls the difference between the Fermi levels of the electron reservoirs at the contacts. 
The differential conductance $dI/dV_{sd}$ is measured by using a standard lock-in excitation of 40 $\mu \text{V}_{\text{rms}}$ at 116 Hz applied to the source electrode. The microwave line consists of a 3.5-mm-diameter beryllium in stainless-steel coaxial line (UT-141) and it is ended by an unmatched dipole antenna. \\
As in similar doped \cite{leti_switching} and undoped \cite{SanquerCornerEffect} devices with tri-gate geometry, the onset of quantum transport takes place via two parallel paths at the corners of the nanowire. Two sets of peaks differing by orders of magnitude in conductivity provide distinct capacitive couplings with the gate (Figs. \ref{fig:dev}a and \ref{fig:dev}b).\\
The highly conductive path $\text{D}_{\text{qd}}$ is attributed to a donor close to the edge of the silicon nanowire with electronic states hybridized with the interface states of the corner. For the exhaustive discussion see Supplemental Material  \footnotemark[1]\footnotetext[1]{see Supplemental Material at [URL] for temperature-dependent measurements and data analysis procedure.}, in agreement with values of donors embedded in etched-silicon channels \cite{Rogge_kondoNL, Rogge_kondoPRL, Sanquer_valleyorbit} Its first two (subthreshold) peaks $D_{\text{qd}}^0$ and $D_{\text{qd}}^-$ are associated to transport via neutral and negative charge states of the phosphorous atom respectively. At higher gate fields the potential barriers at the ends of the tiny channel originate quantum dot-like states visible as above-threshold resonances and Coulomb diamonds with $N \ge 3$ electrons confined. Such donor based quantum dot system is referred as the donor hereafter. 
\section{RESULTS AND DISCUSSION}
\subsection{Valley blockade}
\begin{figure}
\centering
	\includegraphics[width=\columnwidth]{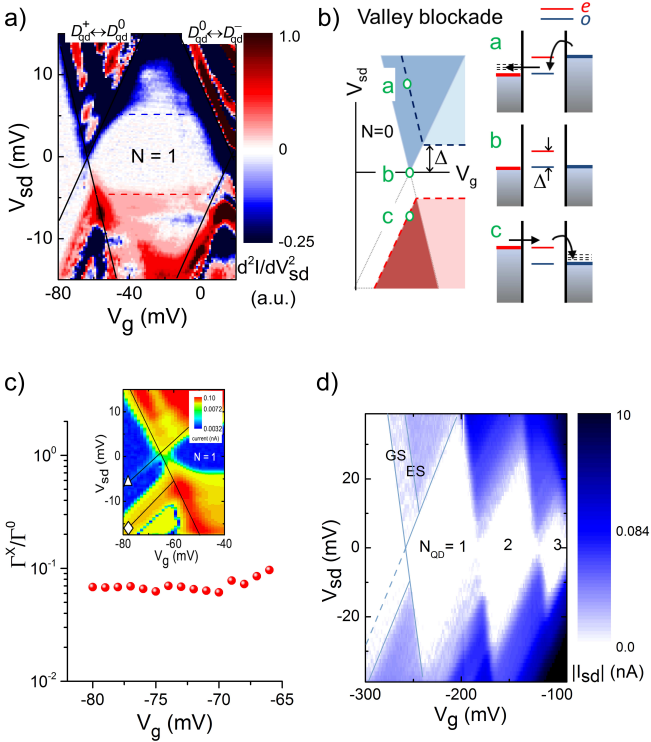} 
	\caption{(Color online) (a) Stability map of the $D_{\text{qd}}^+ \leftrightarrow D_{\text{qd}}^0$ resonance and first Coulomb diamond ($N=1$). Dashed lines indicate the onset of cotunneling via the excited state. For a better visualization the derivative of the differential conductance is computed. (b) Expected stability map in presence of inelastic cotunneling and valley blockade and, on the right, energy diagrams for sequential transitions. Thin dashed lines represent empty states available in the reservoirs. (c) Relative comparison of tunneling rates between states with identical and opposite valley composition. Inset: current stability diagram from which the value $I_{\triangle}$ and $I_{\Diamond}$ used in Equation (\ref{eq:Gammi_ratio}) are obtained. (d) Transport data before the onset of conduction through the donor. The sensitivity is tuned to magnify the current contribution of QD. The first peak shows the asymmetry due to the opposite valley polarization of the source and drain electrodes and the consequent valley blockade. GS and ES denote the ground and the excited state respectively.}
	\label{fig:VB}
\end{figure}

The valley blockade arising at single electron filling is observed by the asymmetry of the stability diagram of $D_{\text{qd}}^+ \leftrightarrow D_{\text{qd}}^0$ transitions in Figure \ref{fig:VB}a. The lack of conduction in the section of the map involving the ground state only at negative bias voltages is the hallmark of valley blockade regime predicted in Ref.\citenum{Prati_valley} as reported in Figure \ref{fig:VB}b. It is determined by perpendicularity of the orbital states at the Fermi energies of the contacts and the ground state of the donor, which is labeled with the odd parity index $o$. The first excited state of the donor, labeled with the even $e$ parity index, defines in addition the inelastic cotunneling region delimited by the horizontal dashed lines.  
The valley splitting $\Delta$ is 6.2 meV \footnotemark[1]. The blockade is lifted at $V_{sd} =-6.2$ mV when the donor excited state is resonant with the $e$ states at the Fermi level in the left lead.\\
The origin of the valley blockade regime is ascribed to the random graining of the As atoms of the contacts in proximity of the active region, which originates singularities in the reservoir density of states \cite{mottonen_DOS}. As the valley parity of the wave functions depends on low dimensionality of randomly diffused dopant distribution \cite{boykin2004}, the electron states at the Fermi levels of source and drain result unintentionally of opposite valley. \\
Such very special condition is confirmed by the stability diagram of the weakly coupled quantum dot in Figure \ref{fig:VB}d. The same asymmetry as the $D_{\text{qd}}^+ \leftrightarrow D_{\text{qd}}^0$ resonance proves from an independent measurement that the valley blockade occurs because of the opposite valley polarization of the contacts.\\
The effectiveness of the valley blockade over the competing Coulomb blockade is quantified through a simple rate equation model: tunneling events between identical valleys turn out to be more than 10 times faster than those between states of different valley. 
The current across a potential barrier, for example the left one, in the sequential regime is $I_{\text{left}}=e (\Gamma_{\text{in}} \Gamma_{\text{out}})/(\Gamma_{\text{in}} + \Gamma_{\text{out}})$, where $\Gamma_{\text{in}}$ and $\Gamma_{\text{out}}$ denote the rates in and out of the donor \cite{bonet_prb}. Any temperature dependence is neglected since we are interested in a ratio between tunneling rates and not in their absolute values.
When $n$ levels of the donor enter in the bias window, $\Gamma_{\text{in}}$ has to be replaced with $\sum_{i=1}^n \Gamma_{\text{in}}^i$ \cite{bonet_prb}. Referring to the band diagrams (b) and (c) of Figure \ref{fig:VB}b, we define $\Gamma_{\text{in}}^X$ as the rate of tunneling events between levels with opposite parity (i.e. from the left lead to the donor ground state).
Analogously, $\Gamma_{\text{in}}^0$ corresponds to tunneling processes preserving the valley index (i.e. from the left lead to the donor excited state).
In particular, along the lines marked by $\triangle$ and $\Diamond$  in the inset in Figure \ref{fig:VB}c, the currents are $I_{\triangle}=e \frac{\Gamma_{\text{in}}^X \Gamma_{\text{out}}}{\Gamma_{\text{in}}^X + \Gamma_{\text{out}}}$ and $I_{\Diamond}=e \frac{( \Gamma_{\text{in}}^0 + \Gamma_{\text{in}}^X) \Gamma_{\text{out}}}{ \Gamma_{\text{in}}^0 + \Gamma_{\text{in}}^X + \Gamma_{\text{out}}}$ respectively.
Since the valley blockade selects only tunneling events entering the donor, relaxations are admitted in the traversing time, giving to a unique outcoming rate $\Gamma_{\text{out}}$.\\
At the stationary state $I_{\text{left}}=I_{\text{right}}\equiv I$.
Because of the valley blockade, we consider $\Gamma_{\text{in}}^X \ll \Gamma_{\text{in}}^0$. By further assuming similar couplings between the donor and the two reservoirs ($\Gamma_{\text{in}}^0 \sim \Gamma_{\text{out}}$), we obtain:
\begin{equation}
\frac{I_{\triangle}}{I_{\Diamond}} \simeq 1 - \frac{1}{1+ 2\Gamma_{\text{in}}^X / \Gamma_{\text{in}}^0}
\label{eq:Gammi_ratio}
\end{equation}
$I_{\triangle}$ and $I_{\Diamond}$ are evaluated from the data plotted in the inset of Figure \ref{fig:VB}c, so that the ratio $\Gamma_{\text{in}}^X / \Gamma_{\text{in}}^0$ is extracted for different values of gate voltage. 
\begin{figure}
\centering
	\includegraphics[width=\columnwidth]{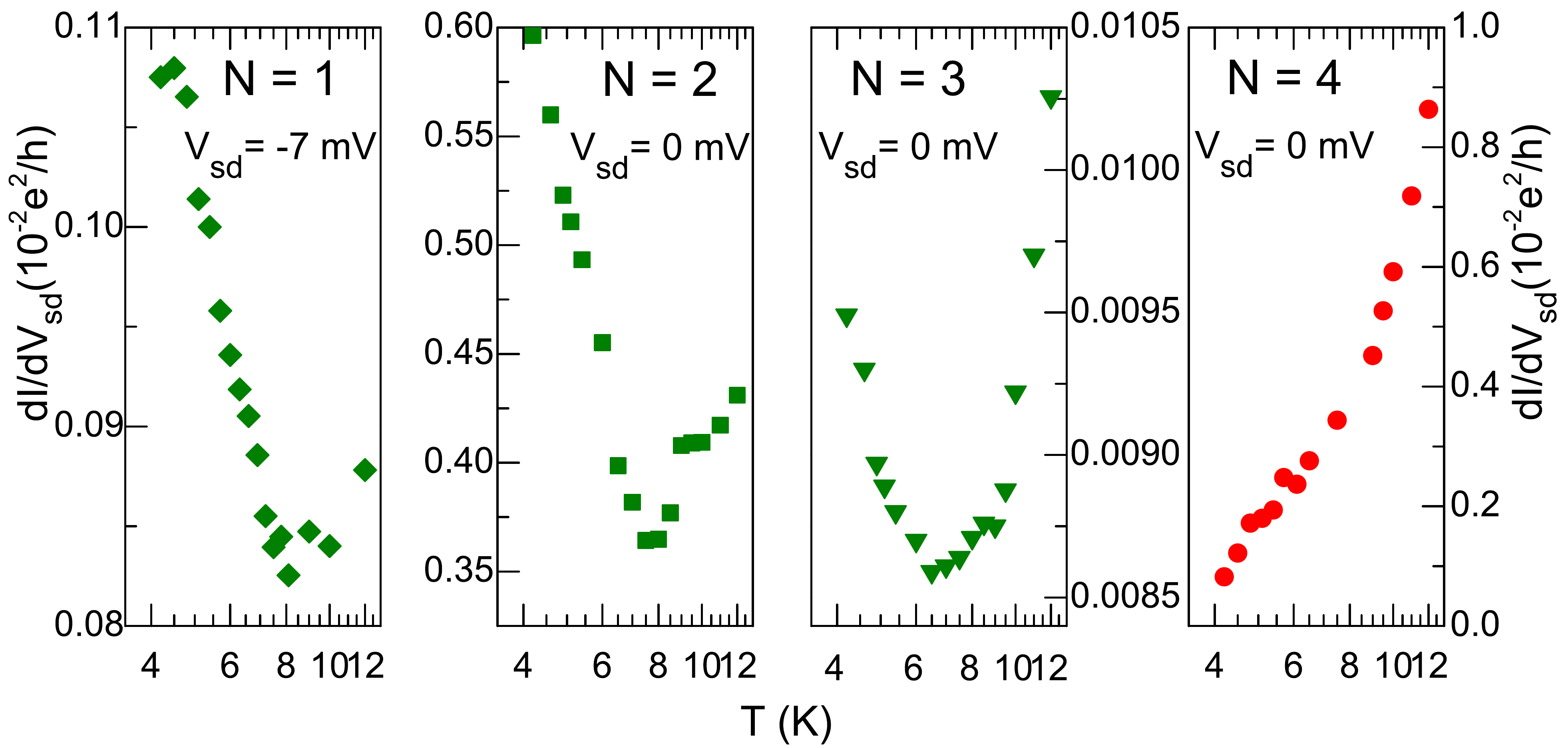} 
	\caption{(Color online) Conductance at different temperatures spanning the first spin-valley shell, for reference see Fig. \ref{fig:dev}b. The increase of conductance at low temperatures is the fingerprint of emerging Kondo physics. Three Kondo temperatures $T_{K1}=(8.2 \pm 1.3)$ K, $T_{K2}= (3.0 \pm 0.4)$ K and $T_{K3}=(3 \pm 1)$ K are obtained at $N=1,2,3$ electron filling respectively. In the Kondo perturbed regime the fitting function adopted is suitable for both SU(2) and SU(4) frameworks \cite{goldhaber_prl_98, Rogge_kondoNL, deFranceschi_kondoCNT}. For $N=4$ the Kondo trend is not observed as the shell is complete.}
	\label{fig:Fig3}
\end{figure}
\subsection{Kondo effect}
We now focus on the experimental investigation of the strong coupling regime to the leads in the first silicon spin-valley shell (i.e. from $N=1$ to $N=4$) of the donor-dot system. In III-V semiconductor quantum dots the presence of the Kondo effect stems from the paired-unpaired spin filling sequence at odd occupation numbers \cite{goldhaber_nat_98, Cronenwett, goldhaber_prl_98}; differently, here the Kondo periodicity is altered by the presence of valley degrees of freedom. A silicon system with an orbital spacing larger than the valley splitting, such as an isolated donor, is predicted to have a Kondo periodicity of four \cite{Shiau2}, as in carbon nanotubes \cite{LPK_spinvalleyblockade}. Figure \ref{fig:Fig3} provides the first experimental signature of a spin-valley Kondo effect reflecting the fourfold degeneracy of the first orbital shell. The hallmark of the Kondo perturbed transport lies in the increase of conductance by lowering the temperature down to 4.2 K at $N=1$, 2 and 3. Such fingerprint is not observed at $N=4$ as no unpaired spin-valley degrees can be screened by the Fermi sea of the leads.
\begin{figure}
\centering
	\includegraphics[width=.8\columnwidth]{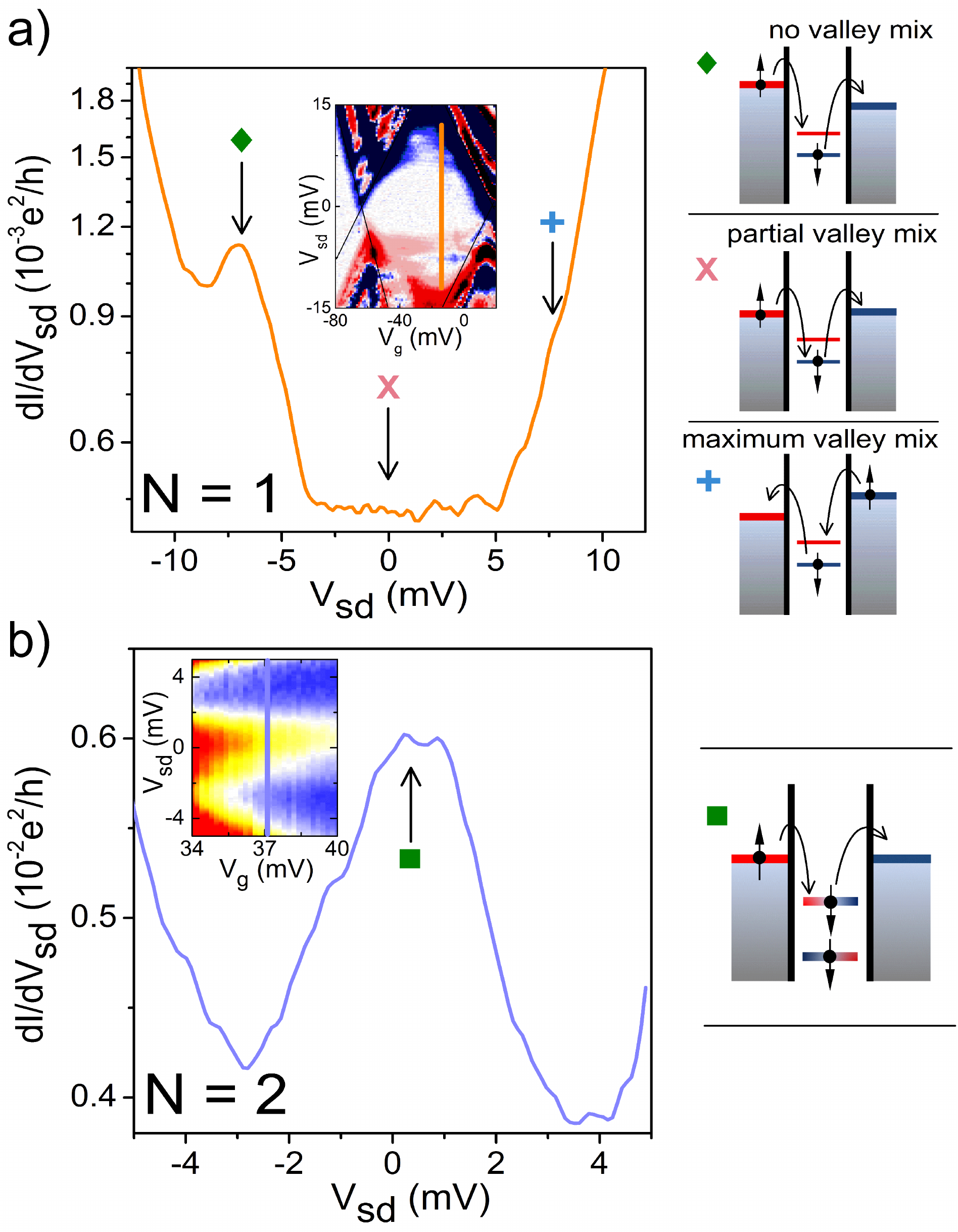}
	\caption {(Color online) Conductance as a function of bias in the $N=1$ and $N=2$ configurations. On the right relative Kondo processes. (a) If the valley parity is conserved the Kondo resonance is observed ($V_{sd}=-7$) mV, whereas at $V_{sd}=0,+7$ mV the Kondo events are prevented by the valley blockade. (b) Schematics describing spin-SU(2) Kondo effect in presence of valley mixing.} 
	\label{fig:Fig4}
\end{figure}
In Figure \ref{fig:Fig4}a the Kondo resonance at $V_{sd}=-7$ mV is marked with a green rhombus. The non-zero conductance background from $|V_{sd}| \gtrsim 5$ mV is attributed to inelastic cotunneling. 
To discern the contribution to conductance due to cotunneling from that Kondo-related, the opposite temperature dependence of the two second-order transport mechanisms is exploited. By heating the sample the Kondo resonance progressively damps till vanishing at 24 K, while the cotunneling background is enhanced \footnotemark[1]. \\
The neutral charge configuration of the donor $D_{\text{qd}}^0$ is shown in detail in the right part of Figure \ref{fig:VB}a. The selection rule on valley parity underlying the valley blockade regime is reflected in second-order Kondo transport: only those processes for which the parity index is conserved clearly stand out from the cotunneling background. Figure \ref{fig:Fig4}a reports a section of the Coulomb diamond. Kondo events with and without valley flips of the electron bound to the donor (expected at $V_{sd}=+7$ mV and 0 mV respectively \cite{Rogge_kondoNL, Shiau1, Shiau2}, marked by black arrows) do not appear. \\
The general Anderson Hamiltonian including valley degrees of freedom reads as \cite{Lim}
\begin{equation}
\label{eq:H}
\begin{split}
& H= \sum_{km\sigma} \varepsilon_k c^{\dagger}_ {km\sigma} c_ {km\sigma} + \sum_{m\sigma} \varepsilon_{m\sigma} d^{\dagger}_ {m\sigma} d_ {m\sigma} \\
& + \sum_{mm'} U_{mm'} n_{m\uparrow} n_{m'\downarrow} + \sum_{km\sigma} V_0 (c^{\dagger}_ {km\sigma}d_{m\sigma} + d^{\dagger}_{m\sigma} c_ {km\sigma}) \\
& + \sum_{km\sigma} V_X (c^{\dagger}_ {k\bar{m}\sigma}d_{m\sigma} + d^{\dagger}_{m\sigma} c_ {k\bar{m}\sigma})
\end{split}
\end{equation}
where $c^{\dagger}, c$ create and annihilate non-interacting fermions in the leads, $\varepsilon_{m\sigma}$ is the single-particle energy level of the donor localized state with valley $m$ and spin $\sigma$, $d^{\dagger}$ and $d$ are the creation and annihilation operators of this state, $U_{mm'}$  the intra ($m=m'$) or inter ($m \ne m'$) valley Coulomb repulsion and $n=d^{\dagger}d$ is the number operator; $c^{\dagger}d$ and $d^{\dagger}c$ describe tunneling events between a spin-valley state in the reservoir and a spin-valley level of the donor and viceversa. 
The channels involving the coupling of the donor with the reservoirs are weighted by $V_0$ for tunneling amplitude of processes preserving the valley index and $V_X$ for exchanges between different valleys ($m$ and $\bar{m}$ indicate opposite parity indices).
\footnotemark[2]\footnotetext[2]{The case of inverse valley polarization of the contacts is not explicitly treated in Equation (\ref{eq:H}): the $c^{\dagger}, c$ operators create and annihilate quasiparticles with both the valley indices since resulting from a canonical transformation which takes into account operators of each reservoir \cite{Glazman88}.}\\
The lack of a zero-bias Kondo peak reveals an intervalley coupling $V_X$ smaller than the intravalley term $V_0$. Kondo processes at $V_{sd}=-7$ mV preserve the electron valley parity: the Hamiltonian in Equation (\ref{eq:H}) reduces to the high symmetric form with just the $V_0$ intravally coupling term. It corresponds to a SU(4)-symmetric Kondo model arising from a twofold spin degeneracy combined with a twofold valley degeneracy; as the cotunneling onset marks a non-zero valley splitting weakly dependent on $V_g$ (Fig. \ref{fig:VB}a), the condition of nearly-degenerate valleys is here achieved by setting a bias equal or larger than $\Delta$.\\
Spin-valley Kondo usually associates with spin-valley blockade, as reported for carbon nanotubes \cite{deFranceschi_kondoCNT, LPK_spinvalleyblockade} where reservoirs and dot are formed within the same tube. Here, as the wave functions in the leads are built from electron states of As with same orbital symmetry as the electron state at the P in the channel, the experiment reveals that in the device under investigation such maintenance of symmetry is achieved thanks to the specific distribution of As atoms in the reservoir areas.
In III-V semiconductors SU(4) symmetric Kondo emerges either from spin-charge entanglement in two electrostatically coupled quantum dots \cite{Goldhaber_entangled} or by inducing multiorbital ground states in vertical dots \cite{sasaki_multilevel}.
In our case, two important clues point towards the SU(4) symmetry: firstly, at operating temperatures of the order of the Kondo temperature $T_{K1}$ the SU(4) physics dominates over the SU(2); secondly, the energy difference between the Kondo peak and the excited state ($\sim 0.8$ meV) is comparable to $T_{K1}$ \cite{Lim}.
The combined emergence of valley blockade and SU(4) Kondo effect leads to a unified picture of quantum transport based on valley parity conservation during tunneling.\\
Differently from $N=1$, the Kondo effect at $N=2$ emerges as a zero bias resonance. 
The participation of the excited state $e$ would imply a Kondo temperature $T_{K2} \sim \Delta/k_B$, where $k_B$ is the Boltzmann constant, yielding $T_{K2} \sim 70$ K, a value that can be easly excluded thanks to the experimental data reported in Figure \ref{fig:Fig3}. Therefore the single-particle level picture has to be replaced with two-electron states obtained from Slater determinants.
Following the formalism previously adopted, the orbital part of such determinants is a combination of $e$ and $o$ states, and the spin component is either a singlet $| S \rangle$ or a triplet $| T \rangle$. Their total antisymmetric tensor products provides the two-particle states: $ | S \rangle| e,e \rangle$, $ | S \rangle| o,o \rangle$, $| T \rangle ( | o,e \rangle - | e,o \rangle)/\sqrt{2}$ and $| S \rangle ( | o,e \rangle + | e,o \rangle )/\sqrt{2}$. 
The electron-electron interaction further mixes the orbital components associated to the three spin singlets. As a result, both singlets and triplets have as orbital counterpart a linear combination of different valley parities (see Appendix).
The Kondo transport takes origin when an electron in the reservoir with a well-defined valley parity tunnels through a barrier by changing its valley index to form the combinations of mixed valley eigenstates above mentioned. For this reason these Kondo exchanges do not suffer any effect of the valley blockade previously discussed: the $V_X$ mixing channel governs the Kondo transitions within the $N=2$ diamond.   
The exchange integral between wave functions with different combinations of valleys is negligible \cite{HE-PRB}: in the limit of non-interacting spins $|S \rangle$ and $|T \rangle$ are degenerate in energy. Hence, each level admits spin flip Kondo processes at zero bias (sketch in Figure \ref{fig:Fig4}b). The ground and the first two-electron excited level are coupled with the conduction electrons of the leads through two orthogonal scattering channels, leading to a two-level SU(2) Kondo effect. That one with higher Kondo temperature is activated here. The vanishing of the exchange interaction hampers a univocal labeling of our Kondo at $N=2$ as under, double-fully or overscreened type.\\
Such an anomalous Kondo effect at even occupancy is inherently related to the non-zero momentum of silicon conduction band edges. In III-V direct-gap semiconductor devices integer total spin states may lead to the appearance of Kondo peaks at even electron fillings. In Ref. \citenum{sasaki_IntegerSpin} a magnetic field brings singlet and triplet into degeneracy, giving rise to singlet-triplet Kondo transitions at zero-bias. Alternatively, by means of an electric field the orbital spacing of two adjacent levels can be tuned \cite{kogan_SingletTriplet} to promote the spin triplet state as the ground state. 
In the system under investigation such condition is experimentally ruled out, as by varying $V_g$ the Kondo resonance remains pinned at zero bias.\\
Within the $N=3$ diamond the contribution to zero-bias conductance due to the Kondo effect (third panel of Fig. \ref{fig:Fig3} and Fig. \ref{fig:Fig4}c) is attributed to the unpaired spin of the third electron fully screened by the many-body spin of the electrons in the reservoirs \cite{goldhaber_nat_98, Cronenwett, goldhaber_prl_98}. \\
\subsection{Microwave suppression of Kondo effect}
\begin{figure}
\centering
	\includegraphics[width=\columnwidth]{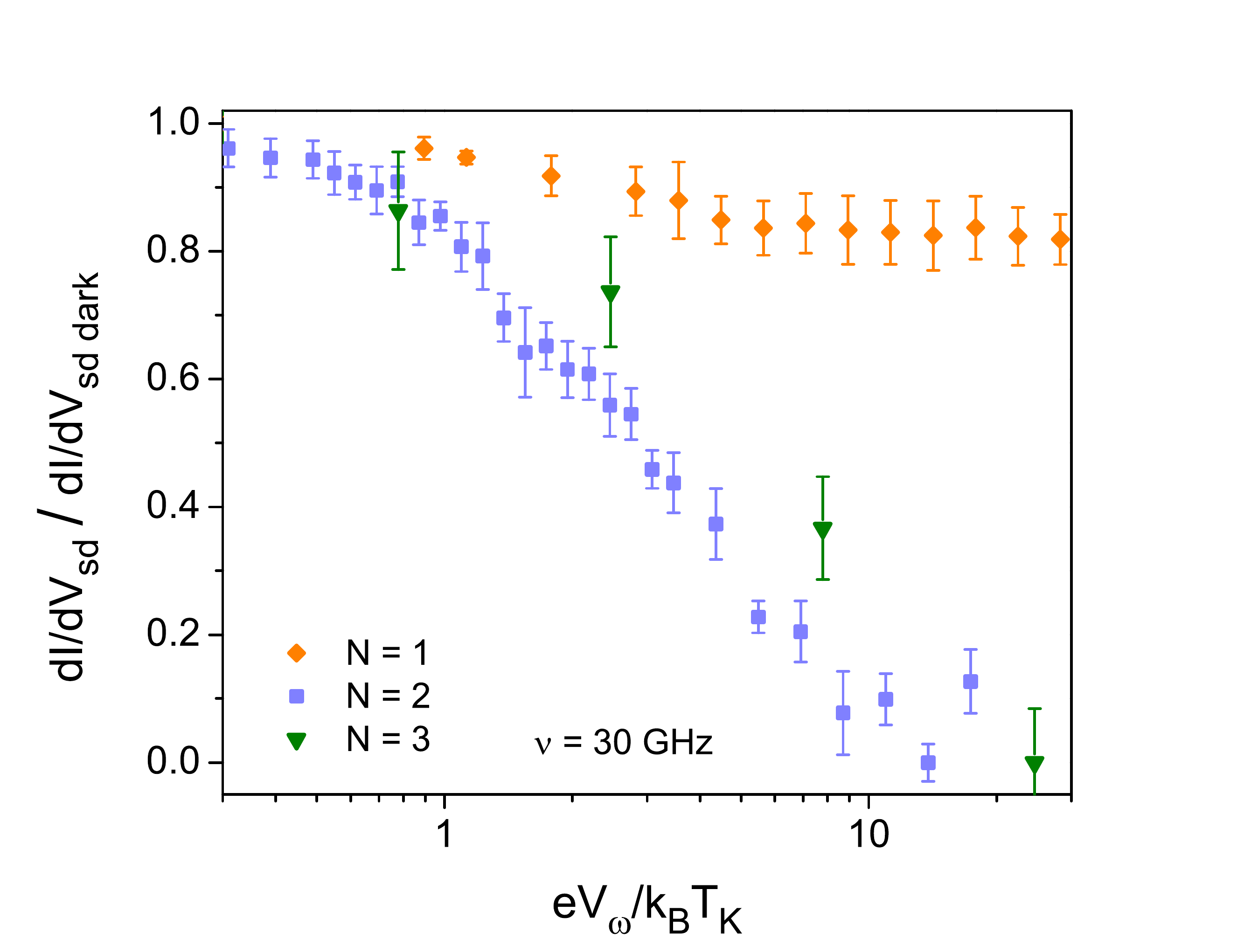} 
	\caption{(Color online) Normalized Kondo peak amplitudes for $N=1,2,3$ as functions of $eV_{\omega}/k_BT_K$ for the three different occupancies.}
	\label{fig:Fig5}
\end{figure}

\begin{figure}
\centering
\includegraphics[width=\columnwidth]{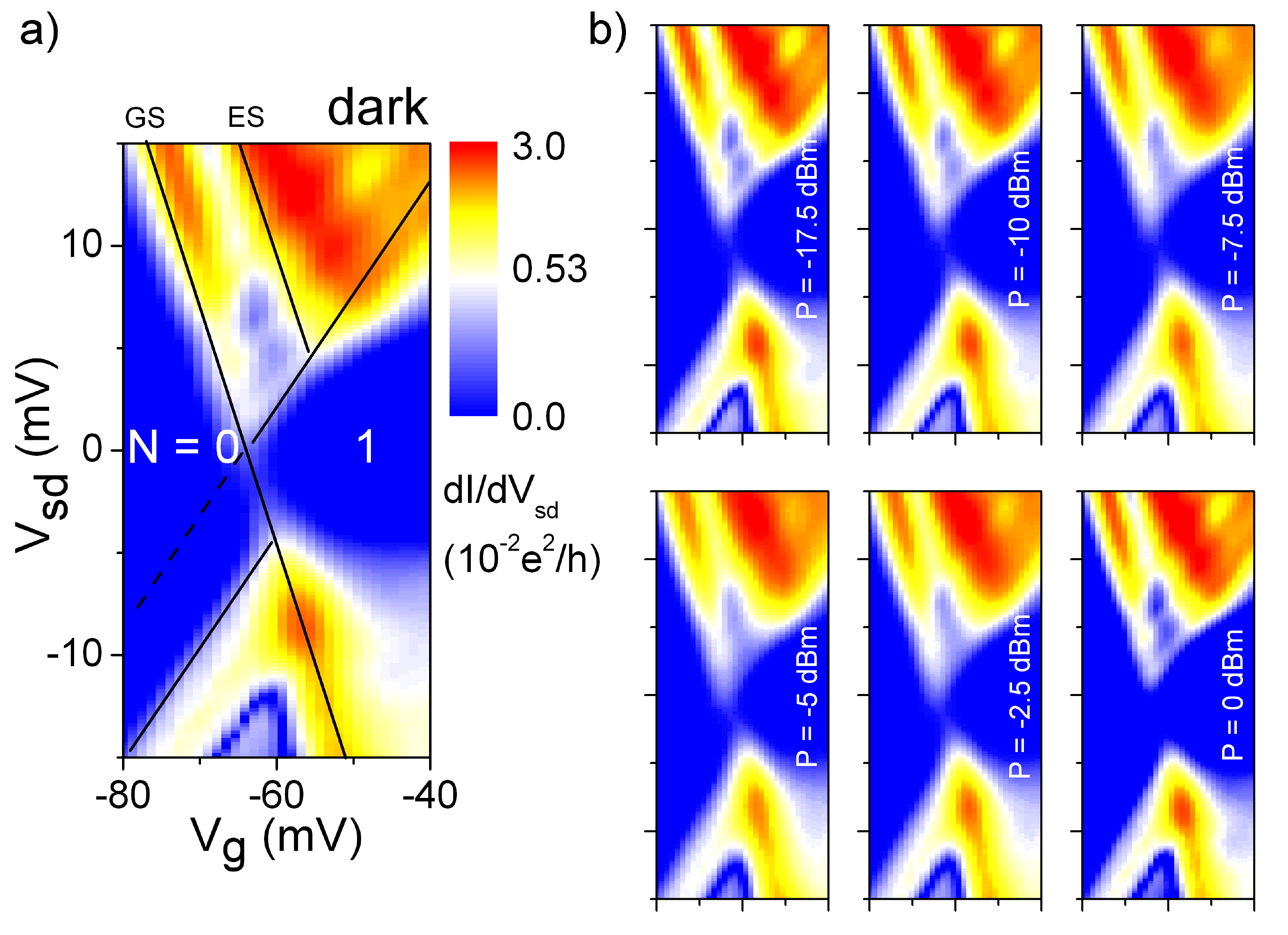}
\caption{(Color online) (a) Differential conductance of the $\text{D}_{\text{qd}}$ resonance without microwave irradiation. (b) Same stability diagram measured at -17.5 dBm, -10 dBm, -7.5 dBm, -5 dBm, -2.5 dBm and 0 dBm nominal powers with a frequency of 30 GHz. The asymmetry due to the valley blockade is not influenced by the microwave irradiation.} 
\label{fig:PP}
\end{figure}

Finally the attention is turned to the perturbation of the system by means of a tunable microwave irradiation. In this section the experimental  microwave suppression of the Kondo effect, previously observed in direct bandgap semiconductors \cite{Elzerman, Kogan, Hemingway}, is experimentally addressed in a multi-valley semiconductor like Si.\\
As the setup is equipped with an unmatched coaxial cable employed as an antenna, the random microwave field distribution induces oscillations both to bias and to donor chemical potential. An empirical conversion between nominal power $P$ and effective amplitude of the perturbation $V_{\omega}$ on the donor site \cite{Elzerman} is established for each electron occupancy ($N=1, 2, 3$) (details reported in Supplemental Material \footnotemark[1]). However, the microwave coupling with the quantum system, as well as the line transmission efficiency, varies also at different frequencies. For this reason, we first developed a robust background subtraction procedure to isolate the Kondo resonances from thermal-activated or photon-assisted background \footnotemark[1]. Next, the Kondo resonance suppression is investigated by varying the frequency. The zero bias resonance at the lowest filling at which it is observed, namely $N=2$ at $V_g=34$ mV, is studied under microwave irradiation at different frequencies.
Above 30 GHz the photon energy approaches $k_B T_{K2}$ and the microwave-induced suppression curves collapse over a single trend \footnotemark[1], revealing that in this regime the microwave field perturbs the coherent spin flip events underlying the Kondo transport.
Because of the high working temperature an appreciable suppression is achieved when many photons cooperate to bring the Kondo state out of equilibrium, $eV_{\omega} > h\nu$. We underline that such suppressive trend stands out independently from the choice of the background fitting function \footnotemark[1]. \\
The systematic comparison among the Kondo resonances at $N=1,2,3$ by varying the power of a 30 GHz excitation is carried out. In Figure \ref{fig:Fig5} the ratio $(dI/dV_{sd})_{\text{peak}}/(dI/dV_{sd})_{\text{dark}}$ is plotted as a function of the parameter $eV_{\omega}/k_BT_K$.
Here "dark" corresponds to the total absence of any external microwave signal.
Such renormalization allows a direct comparison of the three cases $N=1,2,3$ taking into account for the different Kondo temperature.
Microwave irradiation affects the Kondo resonance for all the three occupancies. Such suppression can be ascribed to the quenching effect of the microwave photons rather than to heating, that is excluded by the estimation of the electronic temperature by means of Coulomb blockade thermometry.
Under the assumption of temperature-broadened peak, i.e a cosh$^{-2}$ lineshape, we register an increase of $\sim 0.8$ K at the maximum power (-5 dBm nominally) with respect to the base temperature in dark conditions.
The weak thermal suppression expected in this range of temperatures cannot account for the complete destruction of the Kondo resonance, which has to be attributed to the microwave field.
At amplitudes of energy oscillations smaller or comparable to the Kondo binding energies, such behavior is ascribable to adiabatic variations of the bias combined with weak modulations of the gate potential \cite{Glazman_MW1, Glazman_MW2}, though their separate contribution is difficult to evaluate. By increasing the power of irradiation the rate of photon-induced events increases as well, progressively inhibiting the Kondo processes.\\
The response to microwave signal differs among the cases $N=1,2,3$, as shown in Figure \ref{fig:Fig5}. The slope of the suppression curve for $N=1$ is smaller than that for $N=2,3$, which by contrast are comparable.
An extensive theory \cite{Glazman_MW1, Glazman_MW2} has been developed for a single spin $1/2$ in the Kondo regime (i.e. $T \ll T_K$) perturbed by photons with $h \nu > k_B T_K$. Though a well-established theoretical analysis is still lacking for the mentioned working conditions (multi-valley semiconductor, $h \nu \sim k_B T_K$) we can infer valuable information from our experimental data.\\
The different suppressive rates of Figure \ref{fig:Fig5} suggest a connection between the Kondo symmetries of the three occupancies.
It is notable that the occupancies with similar damping rates have comparable Kondo temperatures and the same underlying symmetry. Such suppressive trends are qualitatively in agreement with previous observations in SU(2) spin Kondo effect \cite{Elzerman, Hemingway}. \\
The reasons of the different rate for $N=1$ are not unambiguously addressed. On the one hand, the microwave suppression could be less efficient on SU(4) than on SU(2) symmetry due to different couplings of the microwave field to spin and to valley degrees. On the other hand, as $T_{K1}>T_{K2},T_{K3}$, stating the different symmetry at $N=1$, the energy carried by 30 GHz photons may result to be less effective in perturbing the Kondo effect because of the higher binding energy $k_B T_{K1}$. \\
Overall, our results show the effectiveness of microwave suppression over the first spin-valley shell, suggesting a possible extension of the theory of Refs. \citenum{Glazman_MW1}, \citenum{Glazman_MW2} beyond the context of standard spin $1/2$ Kondo effect.\\
Finally, we investigate the effect of microwave irradiation on the valley blockade regime. The valley selection rules are not perturbed by the 30 GHz microwave irradiation. In Figure \ref{fig:PP} no significant variations with respect to the dark condition appear in the asymmetric $D_{\text{qd}}^+ \leftrightarrow D_{\text{qd}}^0$ peak by increasing the microwave power. At the Fermi levels of the reservoirs the valley parity electronic states are not altered and the valley blockade phenomenology is fully preserved \cite{Prati_valley}. Such experimental result suggests that the microwave suppression of the Kondo resonance is mainly due to spin coherence breaking, while valley fluctuations remain unaffected.
The Kondo peak at $N=1$ is effectively suppressed at $V_{\omega}=2$ mV \footnotemark[1]. Non-zero peaks of Figure \ref{fig:Fig5} can be therefore ascribed to competing decoherent mechanisms like thermal fluctuations or inelastic cotunneling. 
\\

\section*{CONCLUSIONS}
Valley blockade and multielectron Kondo related transport are observed in a single P donor silicon transistor. The Kondo perturbed regime arises at partial filling of the first spin-valley shell of silicon, $N=1,2,3$. At $N=1$ occupancy the pure valley blockade occuring between donor and leads selects Kondo spin-valley processes at non-zero bias consistent with a SU(4) symmetry. Mixed even-odd valley states originate spin SU(2) Kondo resonances at zero bias when 2 and 3 electrons are confined. The Kondo effect is not observed when the shell is complete at $N=4$, as predicted. Finally, microwave driven oscillations suppress all the three Kondo resonances while leaving the valley blockade phenomenology unaffected.

\section*{Acknowledgements}
The device has been designed and fabricated by the AFSiD Project partners under the coordination of M. Sanquer. The authors acknowledge S. Cocco for technical assistance and M. Belli, A. Chang and D. Culcer for fruitful discussions.

\section*{Appendix}
In this Appendix the spin-valley states are explicitly expressed. In Table \ref{Tab:map} we connect our formalism to the effective-mass theory of Hada and Eto \cite{HE-PRB} by writing the states in both bases. For the two-particle states the superscript 1 labels the first electron wave function and 2 the second electron wave function.\\
By taking into account the Coulomb interaction $\frac{1}{4 \pi \epsilon \epsilon_0 |\textbf{r}_1 - \textbf{r}_2|}$ between the two electrons as a perturbation, the eigenfunctions can be written as linear combination of Slater determinants. The Coulomb interaction is spin independent, so mixed singlet-triplet states annihilate, and just combinations between the three singlets, and the triplet apart, remain. An example of the orbital part of a spin singlet is $\frac{1}{\sqrt{2}}e^{2i\theta} \bigl[- i\sin 2\theta (| e,e \rangle + | o,o \rangle) + \cos 2\theta (| o,e \rangle + | e,o \rangle) \bigr]$, which shows how both even and odd valley parities contribute.
\onecolumngrid

\begin{table}[h]
\begin{tabular}{p{3cm}|c|cc}
\toprule
\rule[-4mm]{4mm}{0cm}
	& Spin & \multicolumn{2}{c} 
{Orbital} \\
\cline{3-4}
\rule[-4mm]{4mm}{0cm}
	& 	& \textbf{Even-odd formalism} & \textbf{Ref. \citenum{HE-PRB} formalism}\\
\hline
\rule[-4mm]{0mm}{1cm}
\multirow{4}{2cm}{Single particle} & \multirow{4}*{ $|\downarrow \rangle$, $|\uparrow \rangle$ } & $|e \rangle$ & $ \frac{1}{\sqrt2} (\psi_z + e^{-i\theta}\psi_{-z})$\\
\rule[-4mm]{0mm}{1cm}
	& 	 & $| o \rangle$ & $ \frac{1}{\sqrt2} (\psi_z - e^{-i\theta}\psi_{-z})$\\
\hline

\rule[-4mm]{0mm}{1cm}
\multirow{12}{3cm}{Two particles} & 	 & $| ee \rangle$ & $ \frac{e^{-i \theta}}{2} ( \psi_{ z}^{(1)} \psi_{-z}^{(2)} +  \psi_{-z}^{(1)} \psi_{z}^{(2)}) +  \frac{1}{2} ( \psi_{ z}^{(1)} \psi_{z}^{(2)} +  \psi_{-z}^{(1)} \psi_{-z}^{(2)} e^{-2i \theta}) $ \\
\rule[-4mm]{0mm}{1cm}
	& $|S \rangle$ & $| oo \rangle$ & $- \frac{e^{-i \theta}}{2} ( \psi_{ z}^{(1)} \psi_{-z}^{(2)} +  \psi_{-z}^{(1)} \psi_{z}^{(2)}) +  \frac{1}{2} ( \psi_{ z}^{(1)} \psi_{z}^{(2)} +  \psi_{-z}^{(1)} \psi_{-z}^{(2)} e^{-2i \theta}) $ \\
\rule[-4mm]{0mm}{1cm}
	& 	 & $\frac{1}{\sqrt 2}  ( | eo \rangle + | oe \rangle )$ & $ \frac{1}{\sqrt 2} ( \psi_{ z}^{(1)} \psi_{z}^{(2)} -  \psi_{-z}^{(1)} \psi_{-z}^{(2)} e^{-2i\theta})$\\
\cline{2-4}

\rule[-4mm]{0mm}{1cm}
	 & $| T_+ \rangle$ & 	 & 	 \\
\rule[-4mm]{0mm}{1cm}
	 & $| T_- \rangle$ & $\frac{1}{\sqrt2} ( | eo \rangle - | oe \rangle )$ & $\frac{ e^{-i\theta}}{\sqrt 2}(\psi_{-z}^{(1)} \psi_{z}^{(2)} - \psi_{z}^{(1)} \psi_{-z}^{(2)})$ \\
\rule[-4mm]{0mm}{1cm}
	 & $| T_0 \rangle$ & 	 &	\\
\hline
\hline

\end{tabular}
\caption{\label{Tab:map} Conversion table between the even-odd formalism adopted in the main text and the formalism of Ref. \citenum{HE-PRB}}.
\end{table}

\section*{SUPPLEMENTAL MATERIAL}

\subsection*{Estimated energies of the P atom}
In Figure 1b of the main text two series of conductance peaks differing by an order of magnitude appear. 
Red-orange (i.e. high-current) diamonds give a unique lever-arm factor $\alpha$, thus a single coupling with the gate, observed up to $N=5$.
The blue diamonds of Figure 1b have a different lever-arm factor. The pair of interlaced diamond patterns with different lever-arm factors suggests a circuital scheme with two conductive paths in parallel ($\text{D}_{\text{qd}}$ and QD) with a negligible inter-conductance \cite{kouwenhoven1997electron}.
The first is associated to a phosphorus atom, labeled as $\text{D}_{\text{qd}}$, while the second to a disorder-assisted quantum dot (QD), like previously observed in similar samples. $\text{D}_{\text{qd}}$ localized states arise from the hybridization of Coulomb potential of a donor, with the gate-induced potential well at the $\text{Si/SiO}_2$ interface \cite{Rogge_starkeffect}.

\begin{figure}[h]
\centering
\includegraphics[scale=0.6]{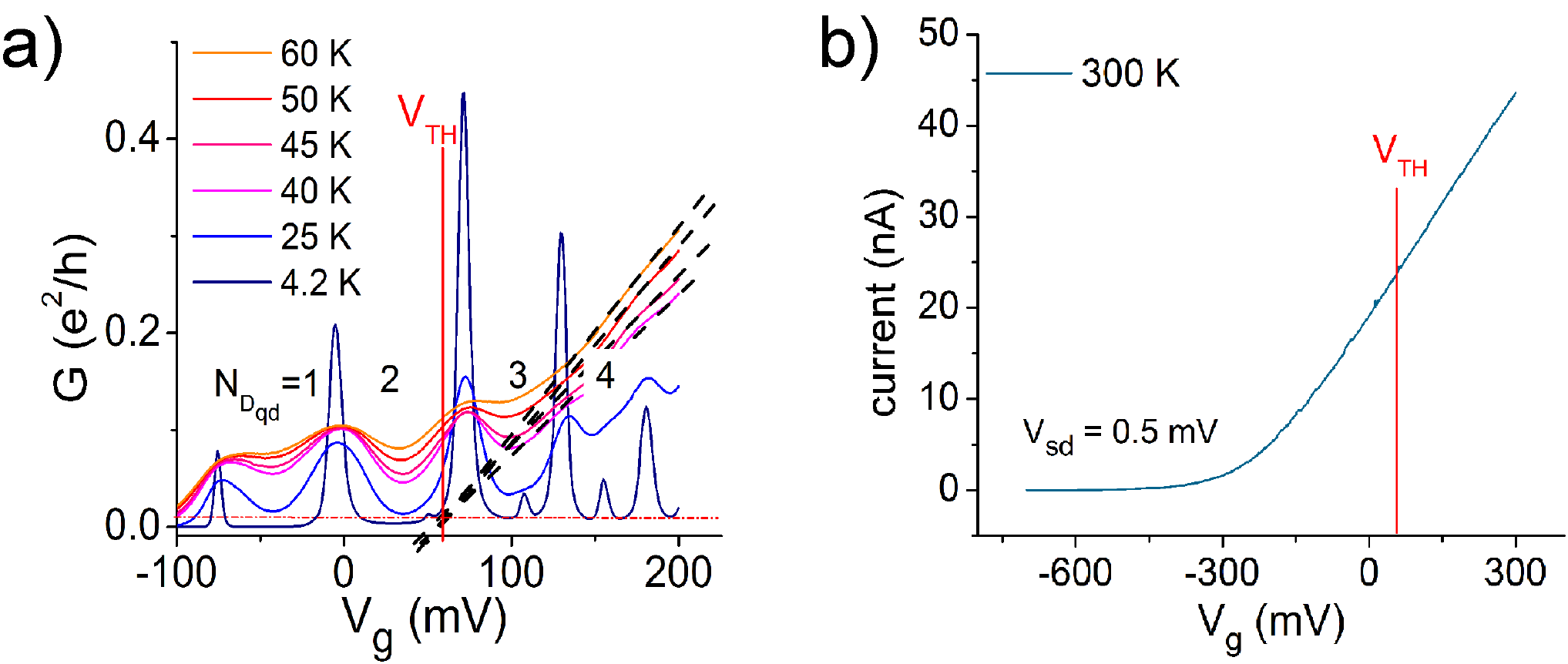}
\caption{(a) The threshold voltage of the $\text{D}_{\text{qd}}$ conductive path is extracted from the linear part of the $G(V_g)$ characteristics at relatively high temperatures. The contribution to $G$ of the QD island is here neglected being one order of magnitude weaker. (b) The presence of donors in the channel is confirmed by the subthreshold current at room temperature.}
\label{fig:threshold} 
\end{figure}
According to the doping concentration, we expect 6-12 donors in the volume of the channel. By lowering temperature thermal transport is progressively quenched so that the threshold voltage $V_{TH}$ can be easly extracted from the linear part of the $G(V_g)$ characteristics, see Figure \ref{fig:threshold}a. At 300 K the subthreshold current is ascribed to thermally broadened tunnelings through the dopants of the active region \cite{wacquez_IEEE} (Fig. \ref{fig:threshold}b). At cryogenic temperatures only those donors sufficiently coupled with both source and drain may be observed from quantum transport. Similarly to previous reports on single atom transistors with etched channels \cite{Sellier, Sanquer_NatNano, Prati_MW}, the first addition energy of $\text{D}_{\text{qd}}$ lies in the range $15-40$ meV (here $U=26.4$ meV). The neutral charge state $D_{\text{qd}}^0$ is $44.7$ meV below the conduction band edge marked by the threshold voltage. This ionization energy is very close to the 45 meV of P atoms in bulk silicon \cite{Kohn}, indicating the proximity of the atom to the gate oxide \cite{wacquez_IEEE}. Such a vicinity enhances the confinement in the gate-field direction, delocalizing the electron wavefunction along the tunneling direction. The consequent strong tunnel coupling with reservoirs allows for cotunneling and Kondo processes. \\
The conclusive argument to attribute the high conductance peaks to a donor is provided by the valley splitting $\Delta$. It is estimated from the $dI/dV_{sd}$ curves as the center of the cotunneling step, see data analysis in Subtraction of the Background section. The value of about 6.2 meV is indeed inconsistent with the typical valley splitting of $\lesssim 1$ meV of a $\text{Si/SiO}_2$ quantum dot \cite{RMP_SiQE}.

\subsection*{Subtraction of the background}
In this section we show that the height of the Kondo peak is irrespective from the choice of the background model. The quantitative analysis of the Kondo contribution to conductance is performed by taking into account both inelastic cotunneling and thermal activated transport. 
By increasing microwave power the Kondo peak amplitude decreases, as a result of oscillating bias and decoherent microwave-induced processes like spin flip cotunneling; on the other hand, the background conductance in Coulomb blocked regions increases due to photon assisted tunnelling. The cotunneling step enhances as well, as predicted by Flensberg \cite{PhysRevB.55.13118} and experimentally confirmed in Refs. \citenum{manscher2003microwave} and \citenum{Ejrnaes}. 
%The same competitive behaviour is observed by increasing the temperature, where a decrease in temperature strenghten the Kondo resonance \cite{Elzerman}, while reducing the cotunneling \cite{manscher2003microwave,Ejrnaes} and background \cite{Elzerman} contributions. 
To discriminate the effects, the data analysis reported in Ref. \citenum{PhysRevLett.92.176801} allows to isolate the Kondo physics from other competitive effects.
\begin{figure}
\centering
\includegraphics[scale=0.6]{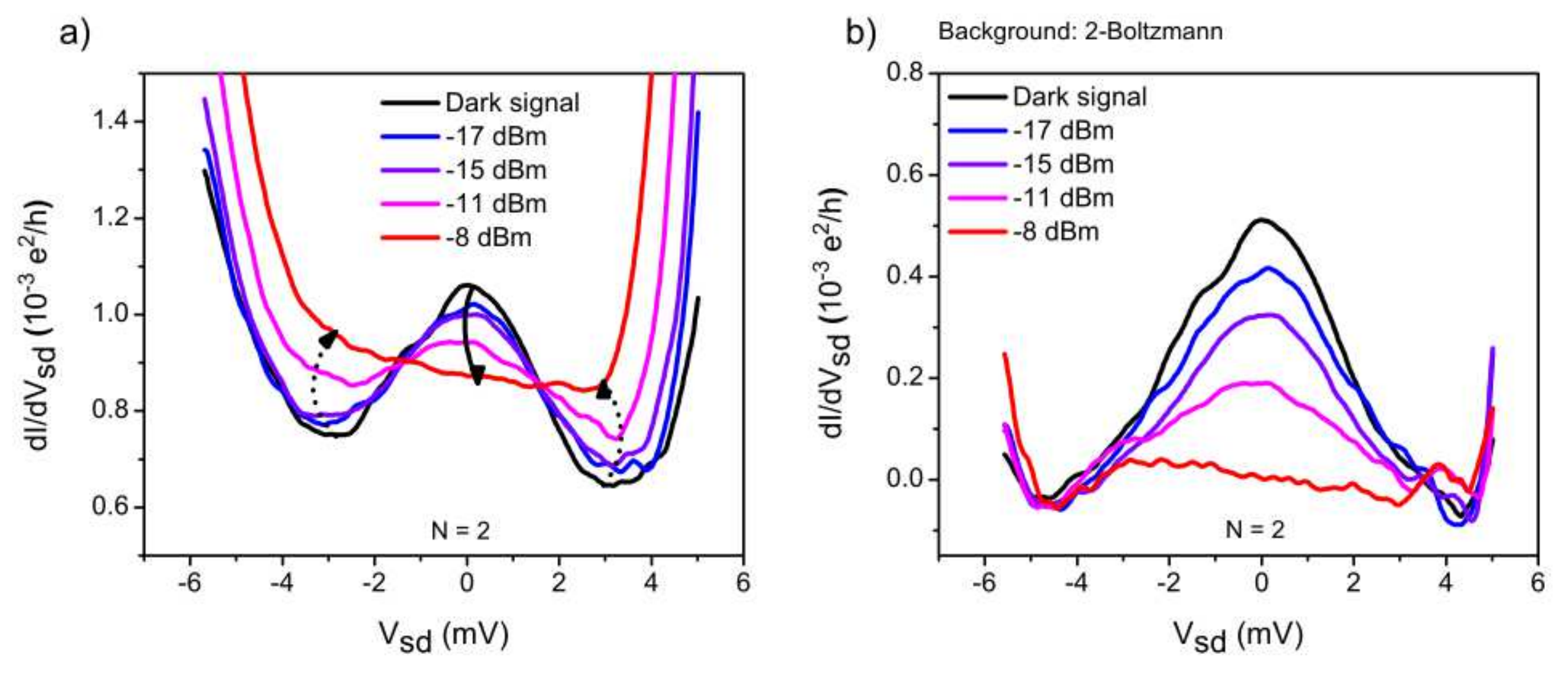}
\caption{(a) Raw data of the Kondo resonance for $N=2$ at different microwave power at 30 GHz frequency. Dashed arrows identify the background enhancement due to photon assisted tunneling, whereas the solid arrow highlights the suppressive trend of the Kondo resonance. (b) Same data after background subtraction.}
\label{fig:N2_resonance}
\end{figure}
\\In Figure \ref{fig:N2_resonance} the zero bias Kondo resonance for $N=2$ is shown at different microwave power values, namely the dark signal and at the nominal applied power -17 dBm, -15 dBm, -11 dBm and -8 dBm, before and after data analysis, with the microwave frequency fixed at 30 GHz. Without any background subtraction at the higher microwave power, i.e. from -11 to -8 dBm, it is difficult to distinguish the Kondo resonance from the Coulomb valley bottom. In Figure \ref{fig:N2_resonance}b the Kondo peak is decoupled from background according to the following procedure. 
We first systematically test three different background functions for $N=1,2,3$. A parabolic, a double exponential and a double Boltzmann fittings are imposed in the proximity of the resonances \cite{PhysRevLett.92.176801}. As an example we report in Fig. \ref{fig:n3} the data analysis for $N=3$. Generally parabola or a double exponential functions return a lower bound for experimental background signal, while the double Boltzmann gives a fit closer to data.
\begin{figure}
\includegraphics[scale=0.45]{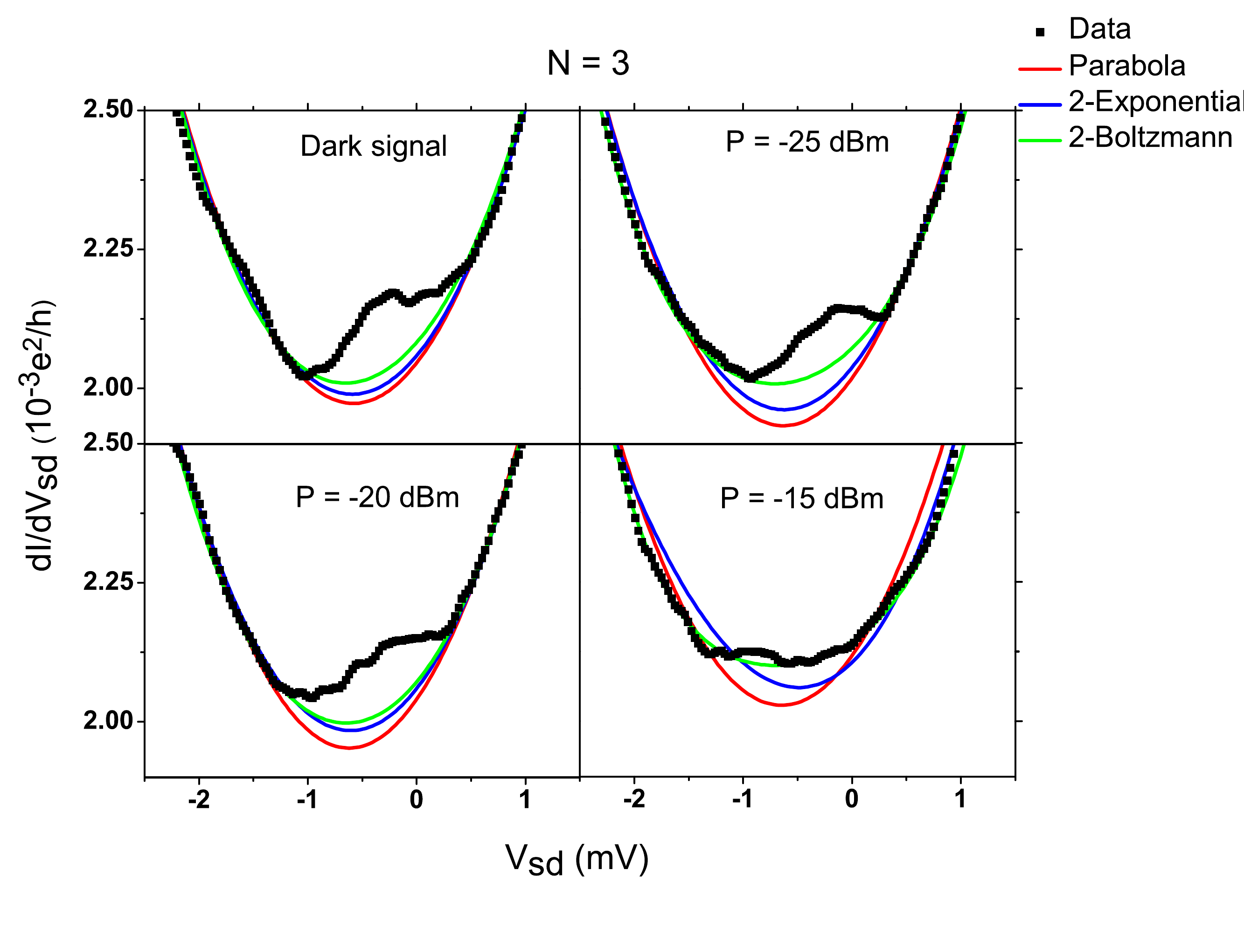}
\caption{Examples of the background fitting using three different functions: a parabola (red), double exponential (blue) and double Boltzmann (green). The data refer to the $N=3$ Coulomb diamond with a 30 GHz radiation except the top-left panel. At  -15 dBm the Kondo resonance is completely suppressed.}
\label{fig:n3}
\end{figure}
Table \ref{Tab:R2} reports the adjusted coefficients R$^2$ obtained by fitting the data of Figure \ref{fig:n3}. It confirms the qualitative conclusions previously derived: the parabolic fit is characterized by the lowest R$^2$ among the tested functional forms, thus signaling the biggest deviation from the experimental data. The double exponential returns a R$^2$ value close, but generally lower, to the double Boltzmann. Analogous results, not shown here, are obtained for $N=1,2$. 
\begin{table}[h]
%\label{Tab:R2}
\centering
\begin{tabular}{ccc}
\hline
Parabola & Double exponential & Double Boltzmann\\
\hline
0.982 & 0.985 & 0.985\\
0.986 & 0.987 & 0.992\\
0.980 & 0.982 & 0.982\\
0.965 & 0.981 & 0.986\\
\hline
\end{tabular}
\caption{\label{Tab:R2}: Adjusted R$^2$ for the fits shown in Fig. \ref{fig:n3}.}
\end{table}
According to these results the data in Figure 5 of the main text and Figs. \ref{fig:N2_resonance}, \ref{fig:lorentz}, \ref{fig:global_fit}, \ref{fig:cotunnelingTOT}, \ref{fig:microw} are extracted after a 2-Boltzmann background subtraction.
\begin{figure}
\includegraphics[scale=0.3]{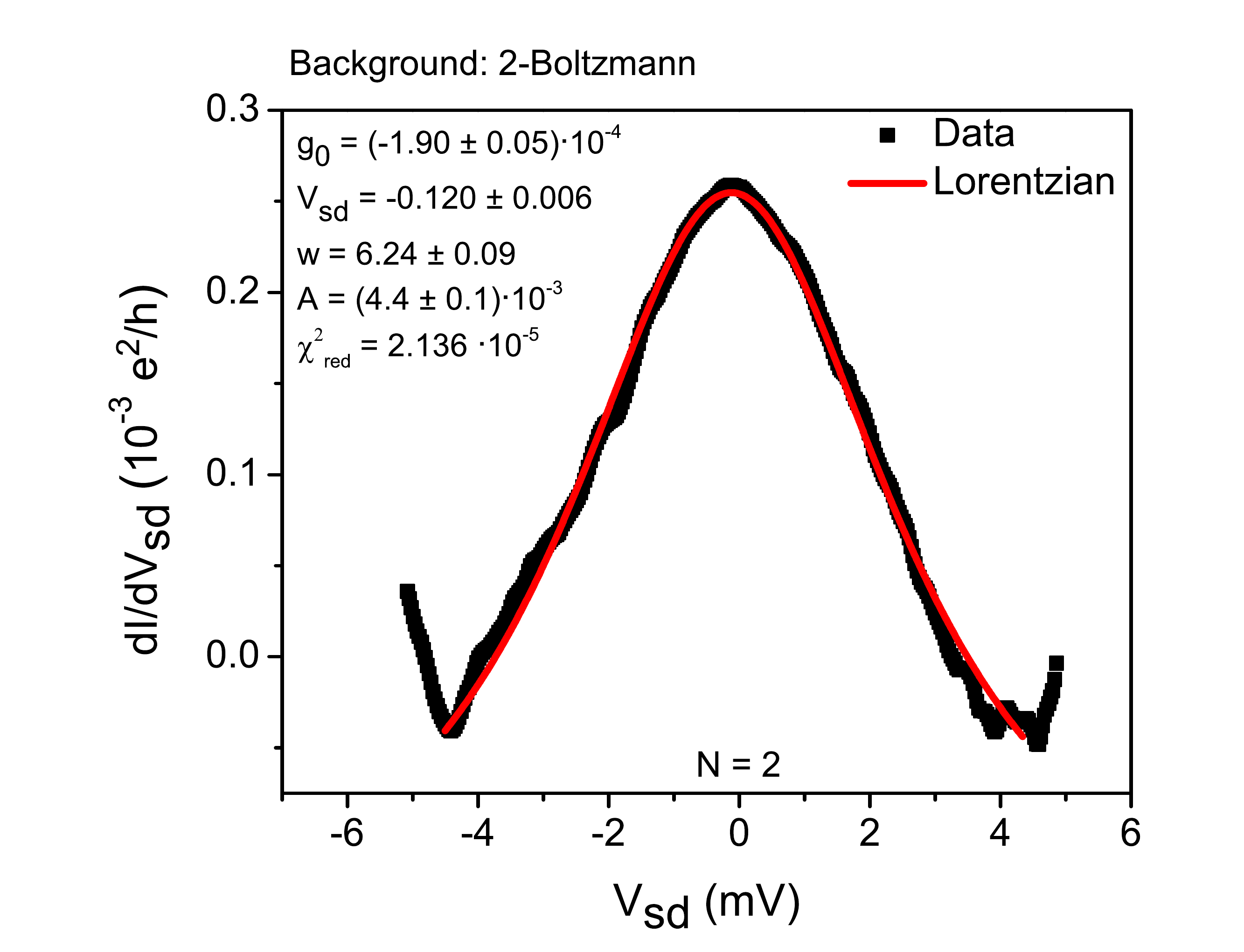}
\caption{Example of a Lorentzian fit performed after background subtraction. The data shown are taken at a nominal microwave power of -19 dBm at 15 GHz, in the $N=2$ valley.}
\label{fig:lorentz}
\end{figure}
Once determined the best functional form of the background signal, the next step is the data analysis of the resulting Kondo resonance. To evaluate the height of such resonances for $N=2,3$ we use a simple Lorentzian (the theoretical shape of the Kondo peak at non-zero temperatures is still debated):
\begin{equation} 
g=g_0+\frac{2A}{\pi}\frac{w}{4(V_{sd}-V_c)^2+w^2}
\end{equation} 
where the fitting parameters are $g_0$, $A$, $w$ and $V_c$. The Lorentzian fits adequately the data, as reported in Fig. \ref{fig:lorentz} where a -19 dBm irradiation is applied at 15 GHz.\\
In order to evaluate and discriminate spin-valley Kondo effect from inelastic cotunneling at $N=1$ we adopt a modified approach.
\begin{figure}
\centering
\includegraphics[scale=0.3]{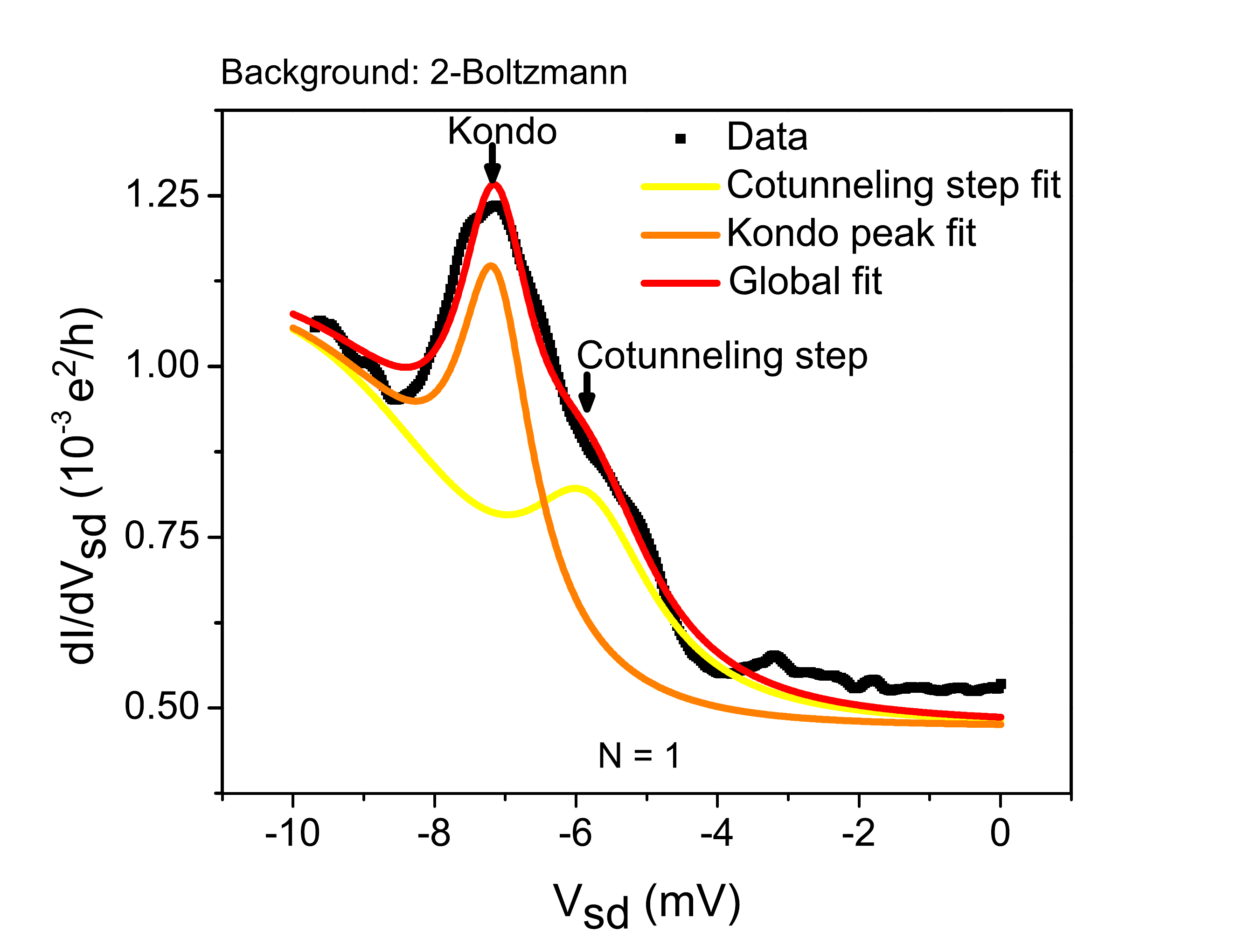}
\caption{Example of data analysis for $N=1$. The non-zero bias resonance is characterized by a non trivial shape, showing a shoulder at -6.2 mV due to the cotunneling step.}
	\label{fig:global_fit}
\end{figure}
The cotunneling signal is predicted to be a step smoothed by temperature, superimposed on a background; in some cases it can be characterized by a cusp-like resonance \cite{wegewijs2001inelastic}. In absence of an overall functional form describing the simultaneous presence of Kondo effect, inelastic cotunneling and background signal we choose to fit our data with the sum of two Lorentzians (one fits the Kondo resonance and one fits the cotunneling resonance) superimposed to each of the three fitting functions mentioned above. At 4.2 K and in dark conditions, the two Lorentzian result centered at $V_{sd}=-6.2$ mV and $V_{sd}=-7$ mV, which are identified as the cotunneling and Kondo peak respectively. The best $R^2$ ($=0.92$) are obtained with the double Boltzmann choice, returning the satisfactory fits of Figure \ref{fig:global_fit}.
\begin{figure}
\centering
\includegraphics[scale=1.1]{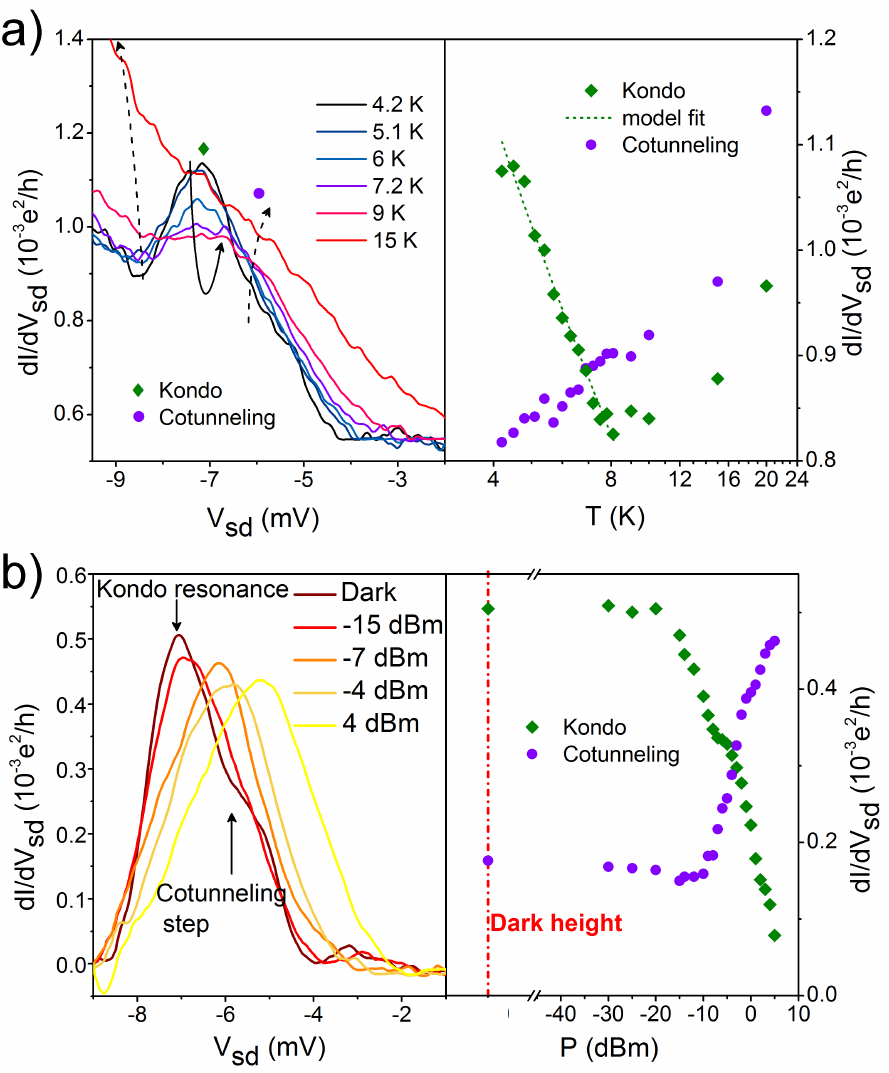}
\caption{(a) Contrary to the Kondo resonance, the cotunneling contribution to conductance is enhanced by lowering the temperature (b) Same experimental data as in Fig. \ref{fig:global_fit} shown after background subtraction at different powers of 30 GHz signal. The Kondo contribution, indeed, decreases whereas the cotunneling one increases resulting in a shift of the maximum of the peak.}
\label{fig:cotunnelingTOT}
\end{figure}
As shown in Fig. \ref{fig:N1fits}, the Kondo resonance height behaviour is insensitive to the chosen background function. It is worthy to note that the fitting procedure with double Lorentzians returns good results when the two resonances are comparable. As soon as one prevails on the other this procedure will never return a complete suppression, producing an artificial saturation of the height of the decreasing peak.
\begin{figure}
\centering
\includegraphics[scale=0.32]{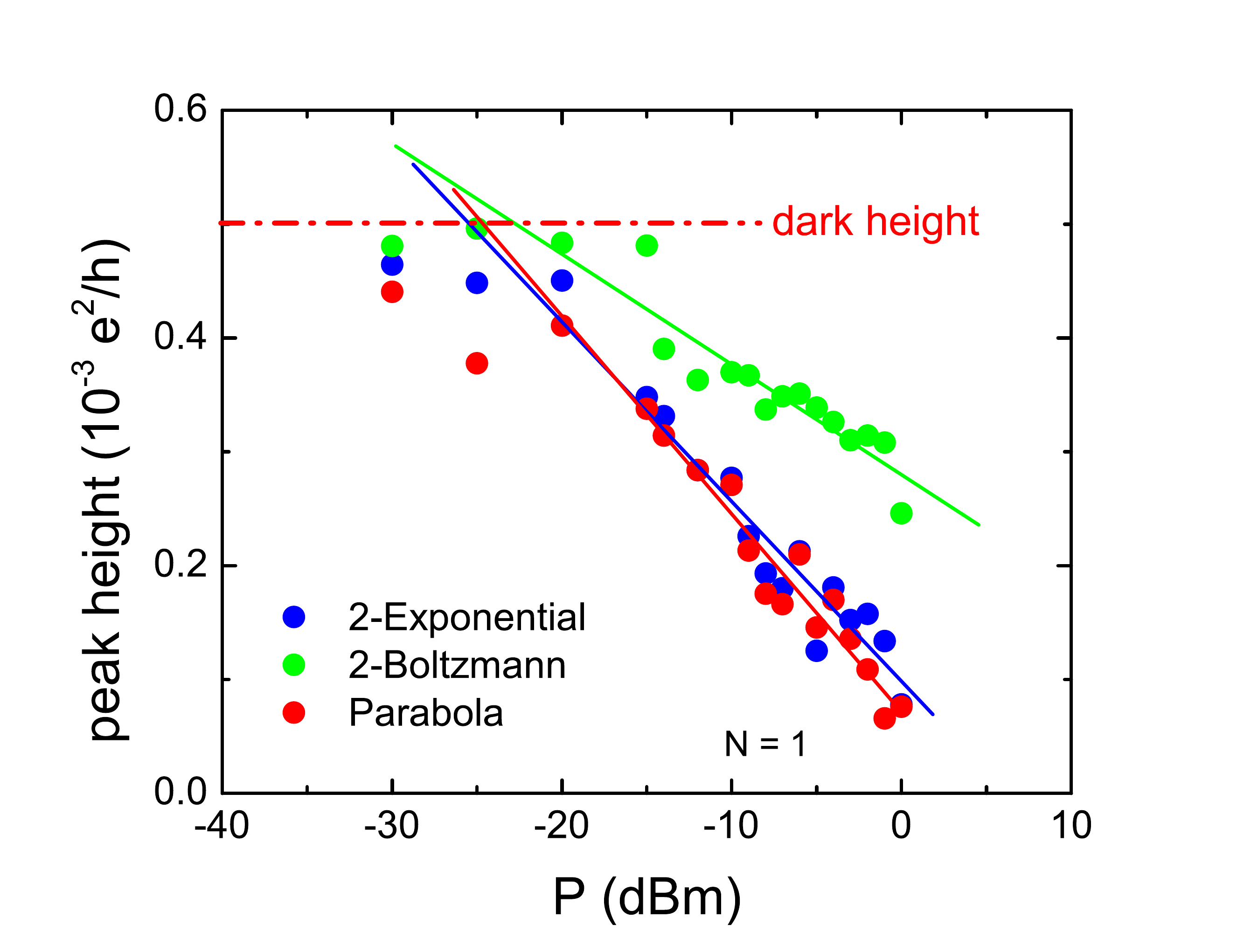}
\caption{Height of Kondo resonance for $N=1$ at 30 GHz as function of nominal power.}
	\label{fig:N1fits}
\end{figure}
\\We finally remark with Figure \ref{fig:cotunnelingTOT}a that the Kondo peak decreases by increasing of microwave power whereas the cotunneling behaves oppositely. The net result consists in a shift of the position of the conductance maximum towards less negative biases, as visible in Figure \ref{fig:cotunnelingTOT}b. 

\subsection*{Rescaling procedure of nominal microwave powers}
\begin{figure}
\includegraphics[scale=0.35]{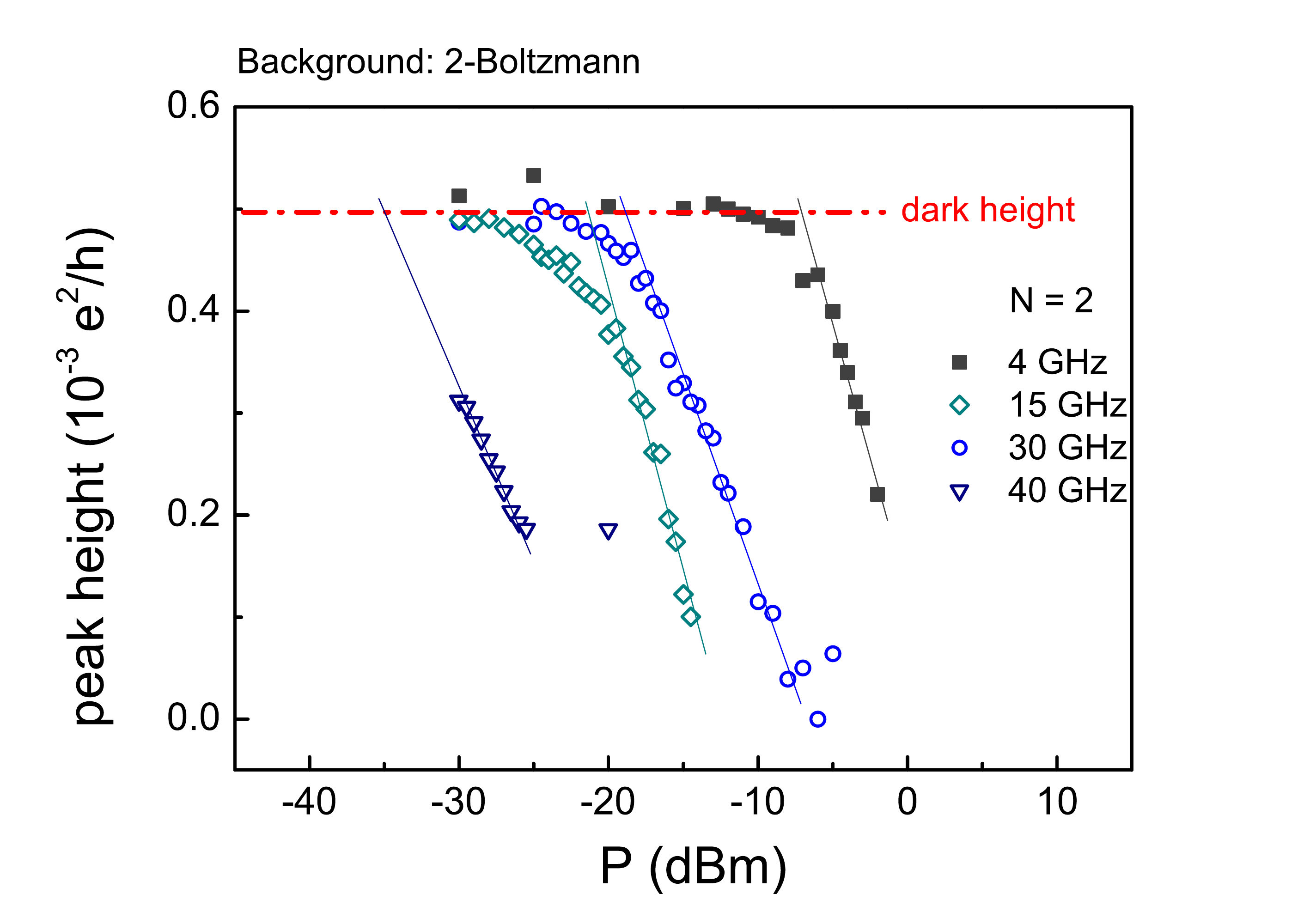}
\caption{Height of Kondo peaks for $N=2$ as function of nominal microwave power at frequencies of 4, 15, 30 and 40 GHz. The power of the onset of suppression $P^*$ is conventionally chosen at the interception between the dark value and the linear decrease of the maxima.}
\label{fig:microw}
\end{figure}

\begin{figure}
\includegraphics[scale=0.45]{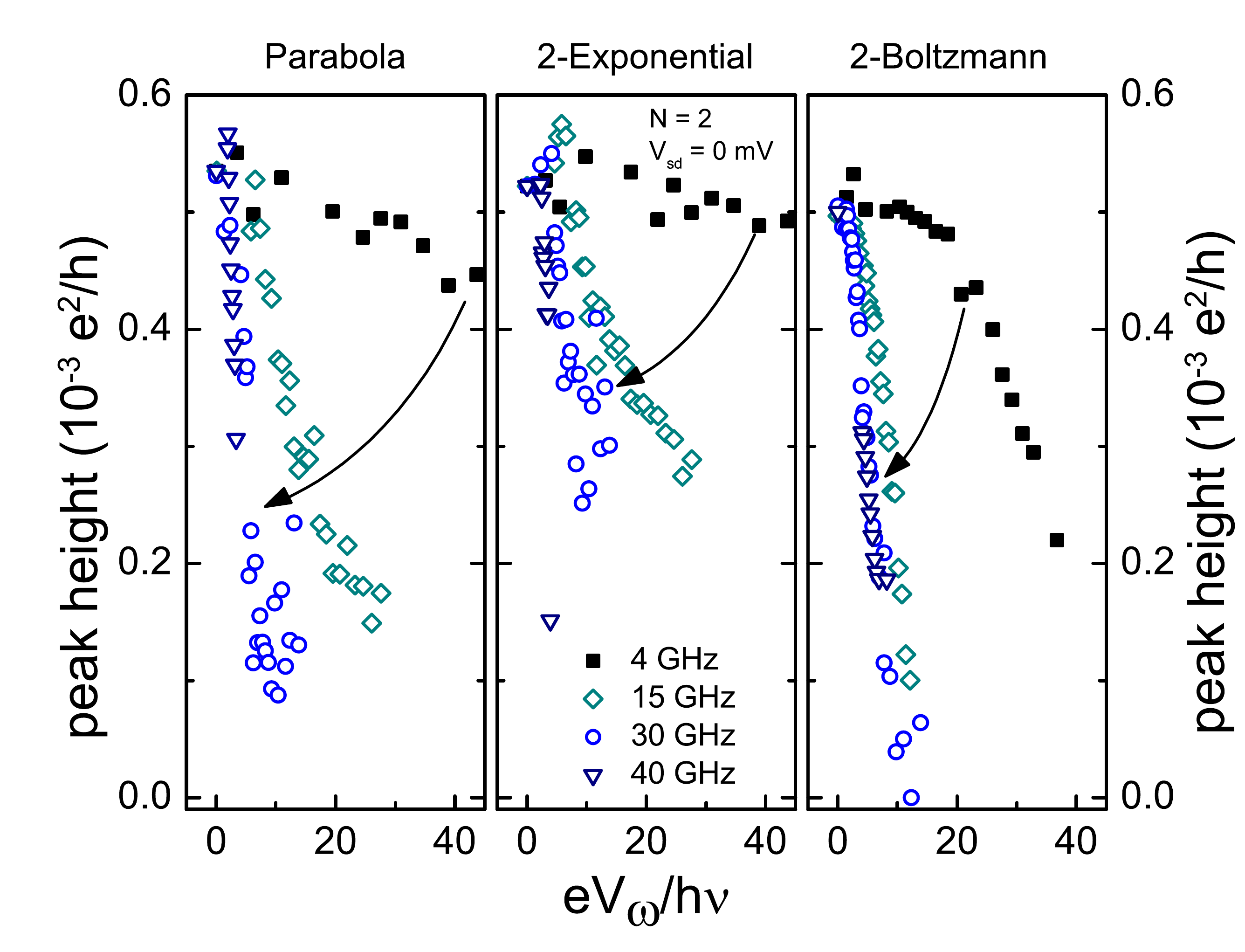}
\caption{Evolution of the Kondo resonance height at different microwave frequency in function of the parameter $eV_{\omega}/h \nu$. In the left panel the subtracted background is fitted with a parabola, in the middle one it is used a double exponential, whereas in the right one a double Boltzmann has been subtracted.}
\label{fig:rescaling}
\end{figure}

As briefly reported in the main text, the transmission efficiency of the ac signal depends on several elements: the nominal 10 dB attenuator, the coaxial cable, the antenna, possible modes excited in the sample holder, behavior of the capping layers of the device. Each element of the line between the generator and the quantum system has its own frequency dependent impedance. We assume that at a fixed frequency the power of the microwave field delivered by the generator is proportional to the square of the amplitude of the signal seen by the quantum system: $P \propto V_{\omega}^2$. In order to determine the constant of proportionality, we follow the procedure devised by Elzerman et al. \cite{Elzerman}: once donoted as $P^*$ the nominal power of irradiation when the suppression of the Kondo peak begins, we set $V_{\omega}^*=k_B T_{\text{eff}}/e$, where $T_{\text{eff}}=5.2$ K from the thermal broadening of the first Coulomb peak at base temperature. Thus, since from our assumption $P^*/{V_{\omega}^*}^2=P/V_{\omega}^2$, we can link each nominal power $P$ to the amplitude of the oscillations $V_{\omega}$ applied to the sample.
Our conventional onset of suppression $P^*$ is extracted by intercepting the dark value of the Kondo resonance with that of the trend when suppression is higher than about 15\%  (10 times the experimental uncertainty), see Fig. \ref{fig:microw}.  
We have checked the robustness of such a procedure by subtracting the three different background functions, thus obtaining different $V_{\omega}$. We have found no significant deviations, which demonstrates that the scaling behaviour observed is independent from our data analysis. Figure \ref{fig:N1fits} is an example for the $N=1$ case.
The evaluation of the amplitude $V_{\omega}$ from the nominal power $P$ allows to replot the heights of the Kondo peaks as a function of the dimensionless parameter $eV_{\omega}/h \nu$.
In Figure \ref{fig:rescaling} we demonstrate that the procedure adopted to calibrate the frequency of the microwave signal is not dependent on the background subtraction.
A comparison of the three panels of Fig. \ref{fig:rescaling} highlights some common features. With any background choice a collapse onto an unique curve is seen for the highest frequencies, i.e. 30 GHz and 40 GHz. At 4 GHz a weak dependence of the peak hight on the parameter $eV_{\omega}/h \nu$ is anyway present. The difference lies in the absolute value of this dimensionless parameter due to the different goodness of the fits. As discussed above we observe the best R$^2$ for the double Boltzmann, reasonable R$^2$ for the double exponential and quite acceptable results for the parabolic signal. The goodness of the fits influences the peak height evaluation, and by consequence the extraction of $eV_{\omega}/h \nu$.  

\subsection*{Microwave effect on a Kondo spin-valley shell}
\begin{figure}
\centering
	\includegraphics[scale=0.75]{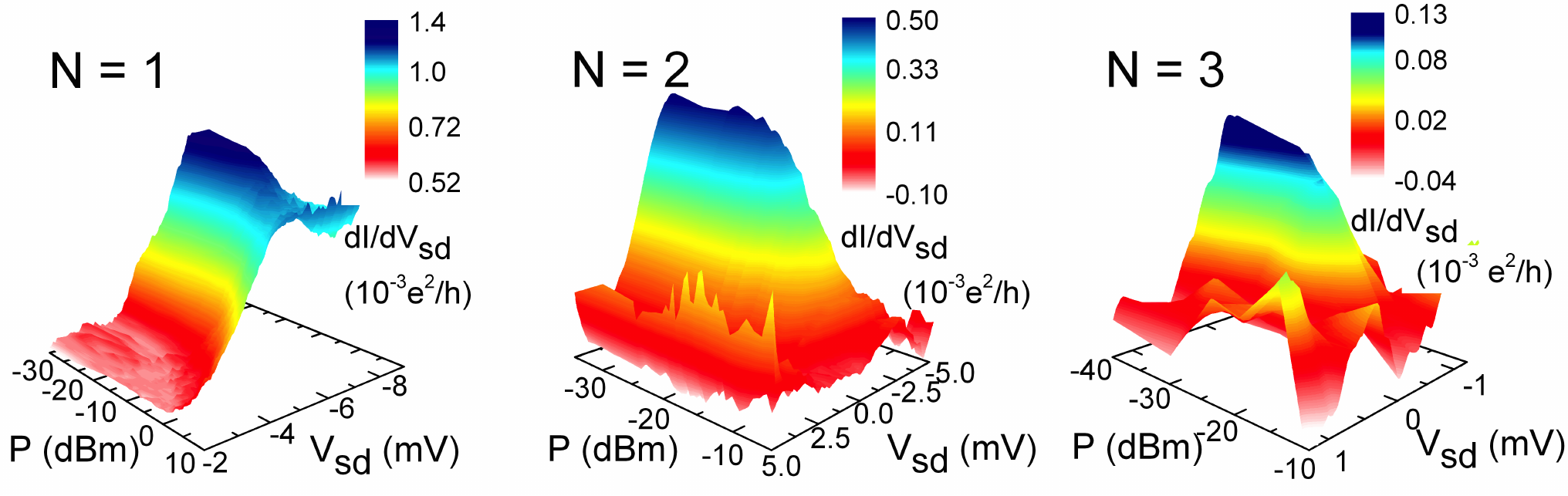} 
	\caption{Kondo resonance suppression as a function of the nominal power $P$ of the 30 GHz microwave field for the first spin-valley shell.}
	\label{fig:3D}
\end{figure}

In Figure \ref{fig:3D} we show an example of the results obtained perturbing the system with a microwave irradiation. Here the responses for $N=1,2,3$ to a 30 GHz radiation are reported. At this frequency the microwave irradiation causes a suppression of the Kondo resonances for the three occupancies.

\twocolumngrid
\bibliography{biblio_GLOBALE}

\end{document}